\documentclass[twocolumn,aps,prl,superscriptaddress]{revtex4-1}

\usepackage[utf8]{inputenc}
\usepackage{graphicx}
\usepackage{multirow}
\usepackage{amsmath,gensymb,amssymb}
\usepackage{siunitx}
\usepackage[bottom]{footmisc}
\usepackage{lipsum}
\usepackage{dcolumn}
\usepackage{bm}
\usepackage{flushend}
\usepackage{xcolor}

\usepackage{hyperref}
\hypersetup{
  colorlinks=true, 
  citecolor=magenta,
  breaklinks=true, 
  urlcolor= blue, 
  linkcolor= blue, 
  bookmarksopen=true, 
}

\usepackage{natbib}

\newcommand{\eff}{_{\mathrm{eff}}}
\newcommand{\ellcap}{\ell_c}

\begin{document}

\title{Experimental Dispersion Relation of Surface Waves Along a Torus of Fluid}
\author{Filip Novkoski}\email[E-mail: ]{filip.novkoski@ens.fr}
\affiliation{Universit\'e de Paris, MSC, UMR 7057 CNRS, F-75013 Paris, France}
\author{Eric Falcon}\email[E-mail: ]{eric.falcon@u-paris.fr}\thanks{corresponding author}
\affiliation{Universit\'e de Paris, MSC, UMR 7057 CNRS, F-75013 Paris, France}
\author{Chi-Tuong Pham}\email[E-mail: ]{pham@upsay.fr}
\affiliation{Universit\'e Paris-Saclay, LISN, UMR 9015 CNRS, F-91405 Orsay, France}

\date{\today}

\begin{abstract}
We report the observation of gravity-capillary waves on a torus of fluid. By means of an original technique, a stable torus is achieved by depositing water on a superhydrophobic groove with a shallow wedge-shaped channel running along its perimeter. Using a spatio-temporal optical measurement, we report the full dispersion relation of azimuthal waves propagating along the inner and outer torus borders, highlighting several branches modeled as varicose, sinuous and sloshing modes. Standing azimuthal waves are also studied leading to polygon-like patterns arising on the two torus borders with a number of sides different when a tunable decoupling of the two interfaces occurs. The quantized nature of the dispersion relation is also evidenced.
\end{abstract}

\maketitle
\paragraph*{Introduction.\textemdash}Vortex rings or toroidal droplets are ubiquitous in nature \cite{SaffmanJFM1978,Marten96,TaddeucciGRL15,WalkerJFM1987,BaumannPoF92,RenardyJFM03,SostareczJFM03,ZoueshtiaghEPJE06,DengPRF17}, but they are unstable. On flat surfaces, they break up into droplets or close their central hole \cite{Worthington1880,PairamPRL2009,BirdNature2010,McGraw2010,MehrabianJFM2013,FragkopoulosPRE2017}. Generating stable tori of fluid thus remains a formidable challenge. It can nevertheless be achieved in toroidal plasma \cite{CothranPRL09,GharibPNAS17}, in biophysics \cite{Simkus2008}, or in fluid mechanics by using an unwetting liquid at the periphery of a cylinder \cite{FalconPRL2019}, by injecting a liquid within a rotating fluid \cite{PairamPRL2009,PairamPNAS2013}, or by levitating a liquid over its vapor film (Leidenfrost effect) on particular substrates \cite{PerrardEPL2012,Ludu2019}.

Due to its periodic boundary condition, a stable torus of fluid is a good experimental system to study the wave propagation in curved and periodic media. For hydrodynamic waves, the curved conditions have been achieved experimentally only for spherical-liquid shells \cite{HoltPRL1996,FalconEPL2009,BerhanuEPL2019}, whereas periodic conditions in planar geometry can be reached (e.g. in an annular water tank \cite{Campbell1953,OnoratoPRL2017}), as well as in curved geometry but without periodicity (e.g. along the border of a liquid cylinder \cite{BourdinPRL2010,PerrardPRE2015}). In this last case, the wave dispersion relation found experimentally \cite{BourdinPRL2010,PerrardPRE2015} and theoretically \cite{Arkhipenko1980,PhamJFM2020,LeDoudicJFM2021} differs from that in planar geometry \cite{lamb}. For a fluid torus, the dispersion relation of waves along the inner and outer torus borders as well as their interaction, are still unreported to our knowledge, the only existing experiment \cite{FalconPRL2019} and the theoretical predictions \cite{FalconPRL2019,Ludu2019} being performed with strong torus constraints.

\begin{figure}[t!]
\includegraphics[width=6cm]{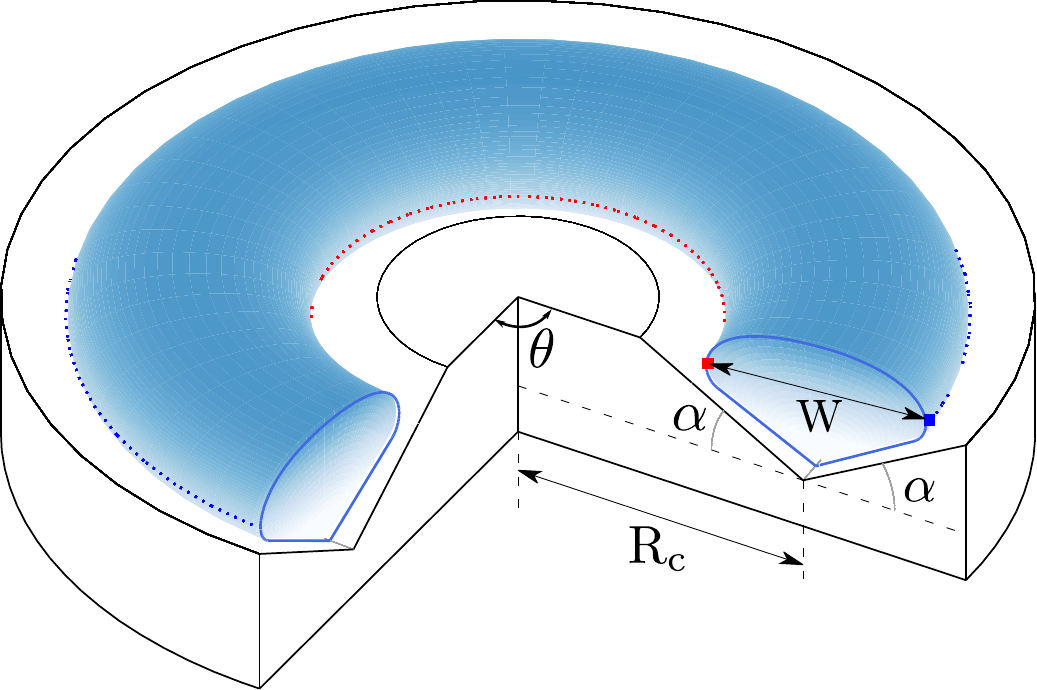}
\includegraphics[width=6cm]{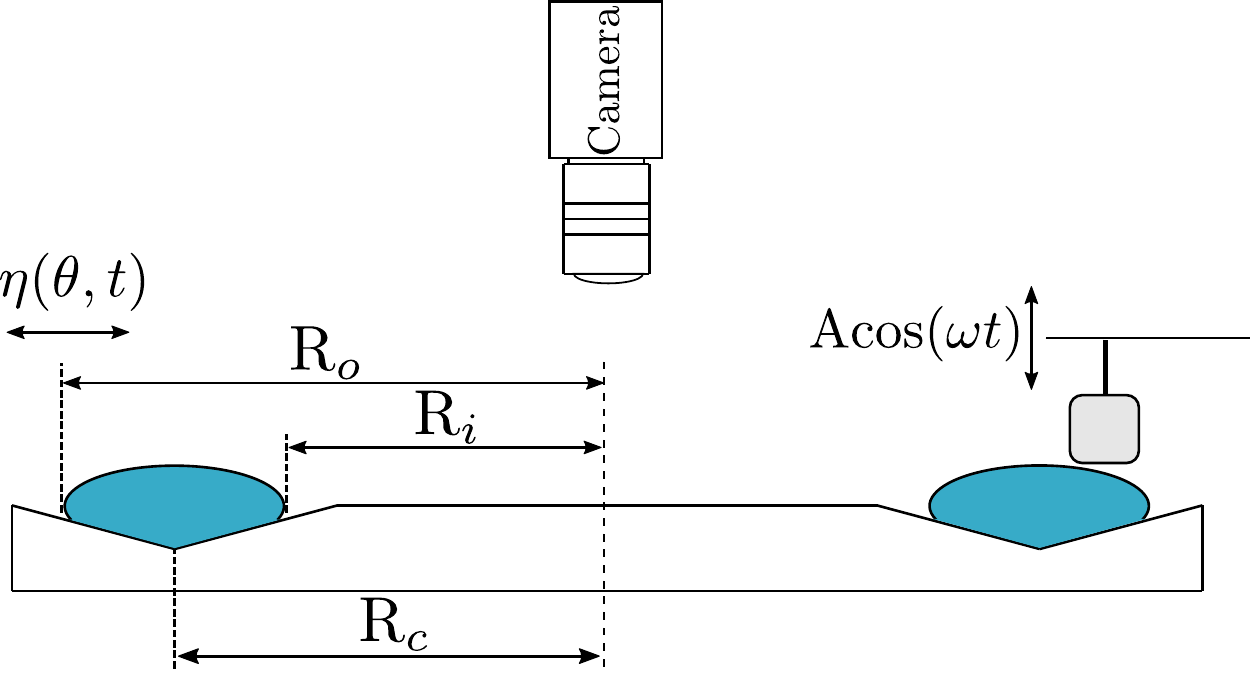}
\caption{Top: Cut-out of the experimental setup showing the water torus on the groove. Bottom: Profile of the setup including the teflon plug connected to a shaker and camera centered above the circular plate.}
\label{fig01}
\end{figure}

Here, we report an original experimental technique to create a stable torus of liquid moving almost completely unconstrained. By means of a simultaneous space and time resolved measurement, we highlight the dispersion relation of azimuthal waves propagating along the torus borders as well as their interaction. Several branches occur with a complex structure reminiscent of forbidden gaps in periodic media in condensed-matter physics \cite{GriffithsAJP2001,Gazalet2013}. Standing waves are also studied showing polygon-like patterns. Our system with these particular boundary conditions (periodic, curved and one-dimensional) including different propagation modes could evidence nonlinear waves, solitons or wave turbulence \cite{FalconDCDS2010,FalconARFM2022} in this geometry.

\begin{figure*}[t!]
\begin{tabular}{cc}
\includegraphics[width=9cm]{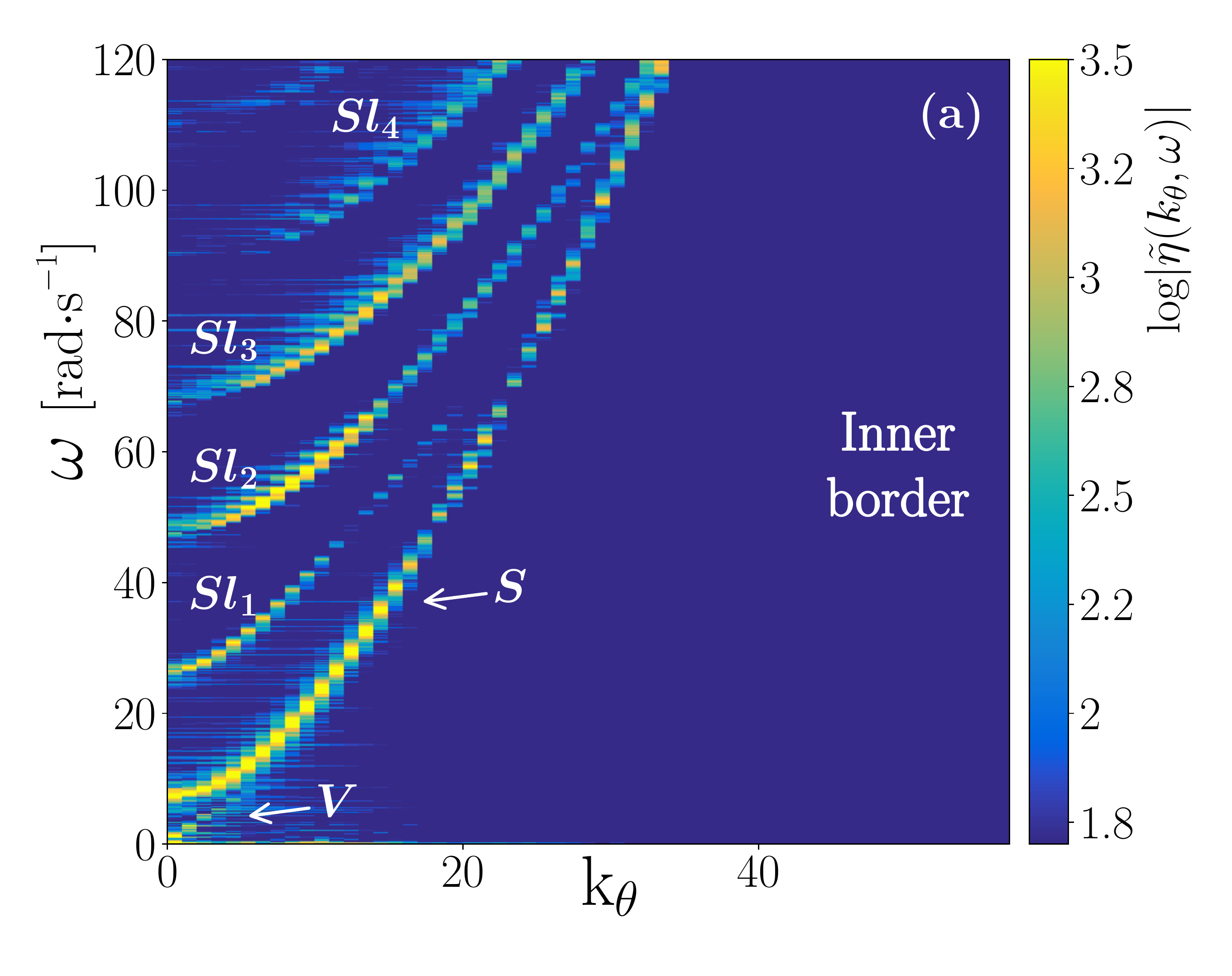} & \includegraphics[width=9cm]{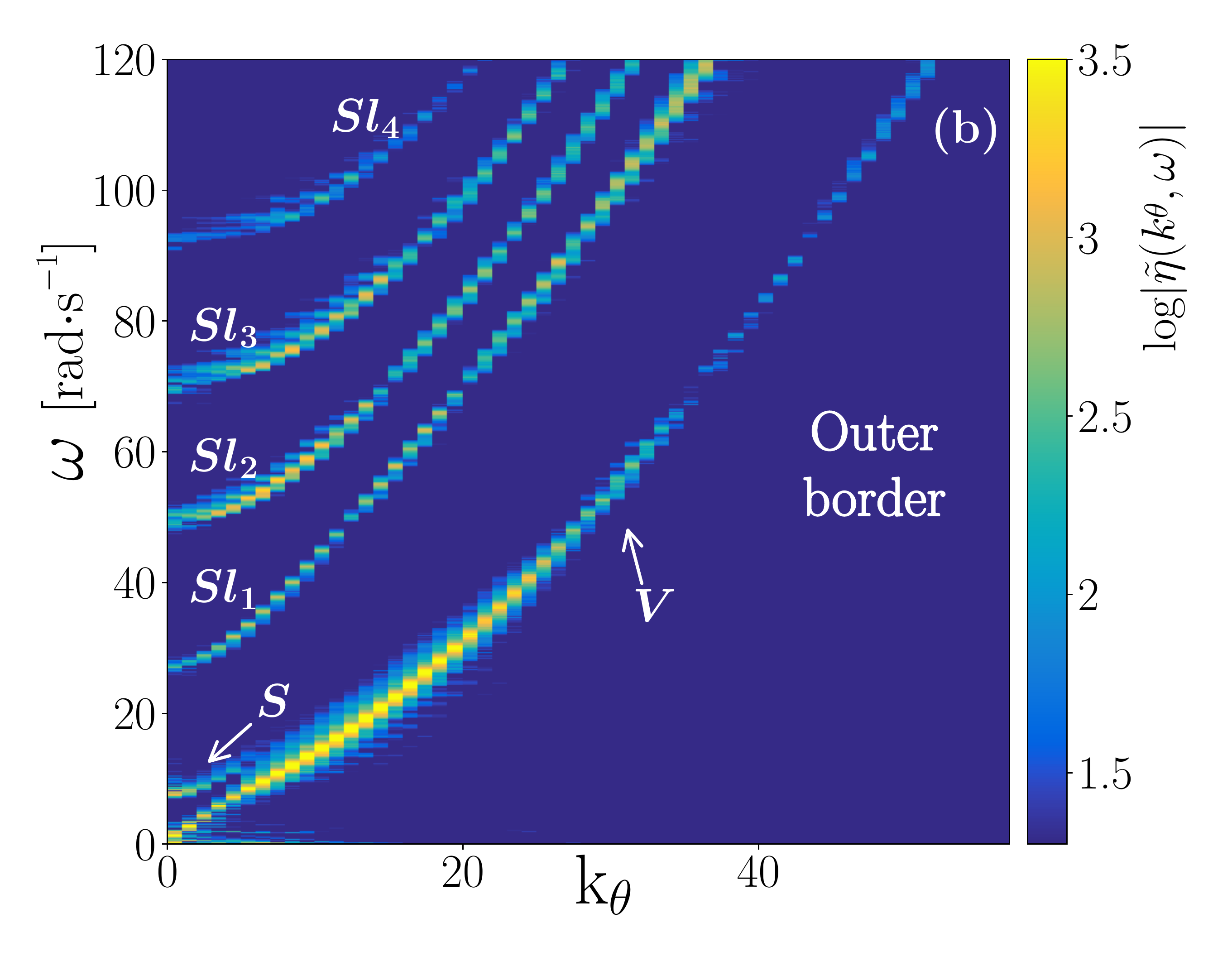}
\end{tabular}
\caption{Dispersion relation $\tilde{\eta}(k_{\theta},\omega)$ of azimuthal waves along the (a) inner and (b) outer borders of a torus of width $W=3.2$~cm, obtained using sweep-sine forcing.  The energy is distributed along distinct branches corresponding to different propagation modes. The \textit{S} branch along the inner border (a) is more pronounced than the \textit{V} branch, in contrast with the outer border (b). The sloshing  \textit{Sl}$_i$ branches are the same for both borders.}
\label{fig02}
\end{figure*}

\paragraph*{Experimental setup.\textemdash}A circular duralumin substrate (19~cm in diameter) is machined to have an axisymmetric wedge-shaped groove, with angle $\alpha=4.5\degree$ to the horizontal, running along its perimeter as shown in Fig.~\ref{fig01}. The substrate is coated with a superhydrophobic spray providing a liquid/substrate contact angle of $160-170\degree$ \cite{neverwet,Gupta2016}. The gentle substrate inclination of angle $\alpha$ constrains the liquid (distilled water) to move along the sloped surface, thus preventing the torus from closing its hole, and experiencing a reduced gravitational acceleration $g\eff=g\sin{\alpha}$ with $g=9.81$~m$\,\cdot\,$s$^{-2}$. 
 
After being deposited on the substrate, we excite the the borders of the torus using a teflon cylinder connected to a shaker oscillating vertically as shown in Fig. \ref{fig01}. For a given fluid volume, a frequency-sweep forcing is applied, i.e., its frequency $f$ is varied linearly in time over $2$~min from $0$ to $20$~Hz. The typical amplitude of vibrations is 3~mm. A camera located above the torus records its interface displacements. Using a border detection algorithm written using \textit{OpenCV}\cite{github}, we extract from the video the horizontal displacements, $\eta(\theta,t)$, of the two borders with respect to their stationary positions. The procedure is repeated by adding water to change the width, $W$, of the torus in the range $1.2$--$3.2$ cm, the center of the section remaining at the groove radius $R_c=7$ cm. These values correspond to torus aspect ratios $\xi=(2R_c/W)$ from $4.3$ to $11.7$. For each volume, we perform a 2D fast Fourier transform on the inner- and outer-border signals, moving from real space $\eta(\theta,t)$ to Fourier space $\tilde{\eta}(k_\theta,\omega)$. This leads to the dispersion relation, $\omega(k_\theta)$, of the azimuthal waves, $\omega$  being their angular frequency and $k_\theta$ their azimuthal wave number, that takes integer values $n$.

\paragraph*{Dispersion relation.\textemdash}As the torus is excited, linear waves propagate along its borders (see video in Supp. Mat. \cite{SuppMat}). We observe that the injected energy is redistributed in different regions of the Fourier space, localized around several branches (see Fig. \ref{fig02}a for the inner, and Fig. \ref{fig02}b for the outer border).
%
The dimensionless wave number $k_\theta$ is used to compare the wave spectra along both borders. It takes discrete values because the system is periodic. We will set $R_i=R_c-W/2$ and $R_o=R_c+W/2$ (see Fig.~\ref{fig01}).

\begin{figure}[t]
\includegraphics[width=\columnwidth]{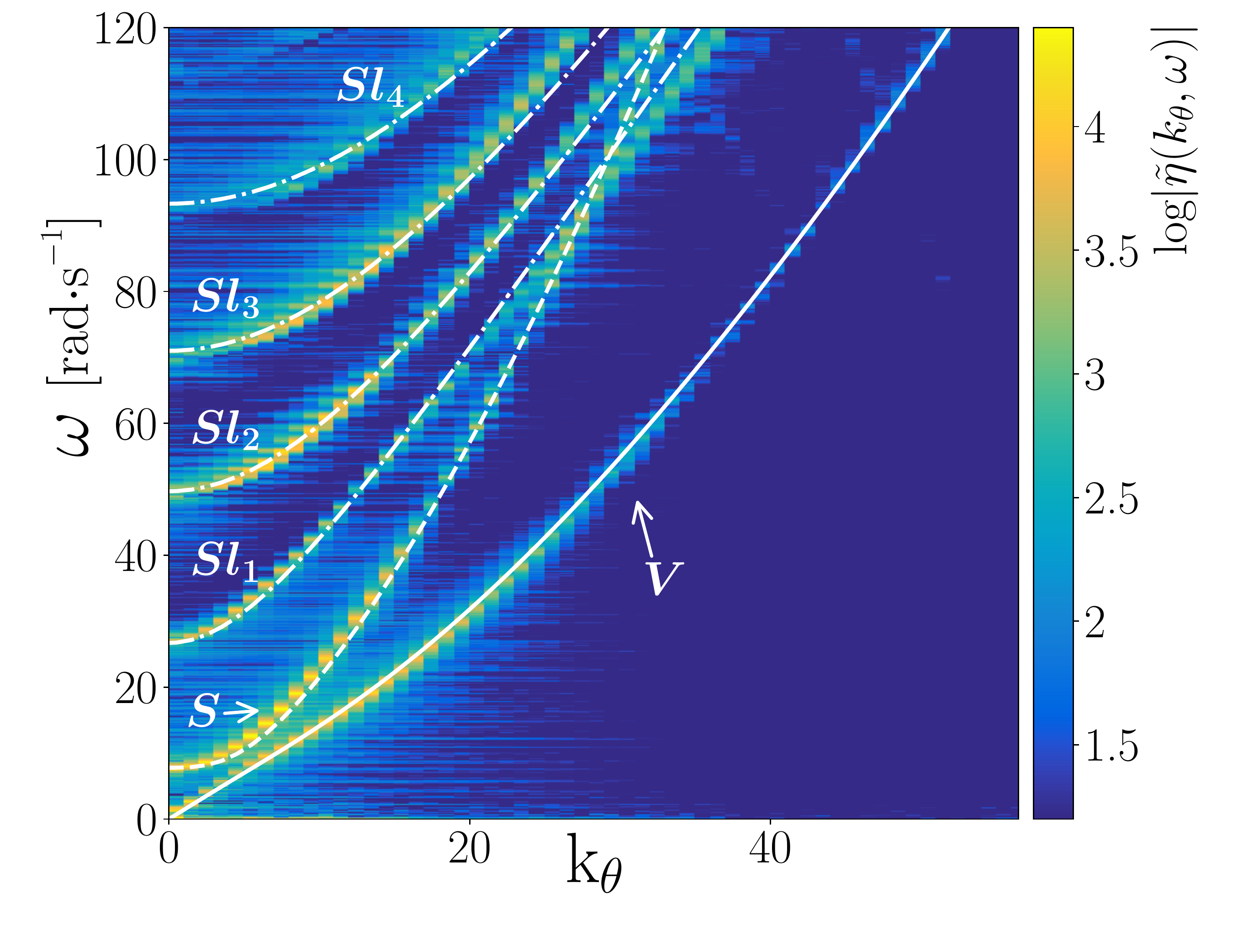}
\caption{Superimposition of the two spectra of Fig. \ref{fig02}, as well as the predicted branches. \textit{V} branch corresponds to Eq.~\eqref{eq:dispersion-relation} (solid line), \textit{S} branch to Eqs.~\eqref{Smode} and \eqref{eq:cutoff}  (dashed line), and (\textit{Sl$_i$}) branches to Eq.~\eqref{Slmode} (dot-dashed lines). $W= 3.2$~cm.}
\label{fig03}
\end{figure}

The different branches observed in Fig.~\ref{fig02} correspond to different modes of wave propagation. The branch passing through
zero, denoted \textit{V} and visible in both spectra, is the dispersive branch stemming from the usual gravity-capillary waves (see below).  The first branch starting above zero, marked \textit{S}, has a cutoff frequency $\omega^{S}_0\equiv \omega^{S}(k_{\theta}=0)$ which depends on the torus volume (see below). Other branches (denoted \textit{Sl}$_i$ with $i=1$, 2, 3 or 4) correspond to sloshing modes clearly different from the \textit{V} and \textit{S} branches. In Fig. \ref{fig03}, we superimpose the two spectra of Fig. \ref{fig02} to better understand their structures. All these modes are described below in detail.

\paragraph*{V mode.\textemdash}The \textit{V} branch is well described by a dispersion relation of gravity-capillary waves  (see solid line in Fig. \ref{fig03}) of the form
\begin{align}\label{eq:dispersion-relation}
  \omega_{V}^2=\left(g\eff\frac{k_\theta}{R_o}+\frac{\sigma\eff}{\rho}\frac{k_\theta^3}{R_o^3}\right)\tanh{\left(\frac{k_\theta W R_o}{2R_c^2}\right)}
\end{align}
with $\rho=1000$ kg/m$^{3}$ the fluid density, $g\eff=g\sin{\alpha} \simeq 0.77$ m\,s$^{-2}$, and $\sigma\eff=60$ mN\,$ $m$^{-1}$, an effective surface tension inferred by fitting the data (regardless of $W\in[1.6,3.2]~$cm); $g\eff$ and $\sigma\eff$ are notably linked to the geometry of the substrate \cite{LeDoudicJFM2021}. For $k_\theta\rightarrow 0$, Eq.~\eqref{eq:dispersion-relation} reads $\omega_{V}=\Omega_\varphi k_\theta$. The angular phase velocity of gravity waves $\Omega_\varphi$ is inferred by fitting the \textit{V} branch in Figs.~\ref{fig02} or \ref{fig03} near $k_\theta=0$. 

\begin{figure}[t]
\includegraphics[width=8cm]{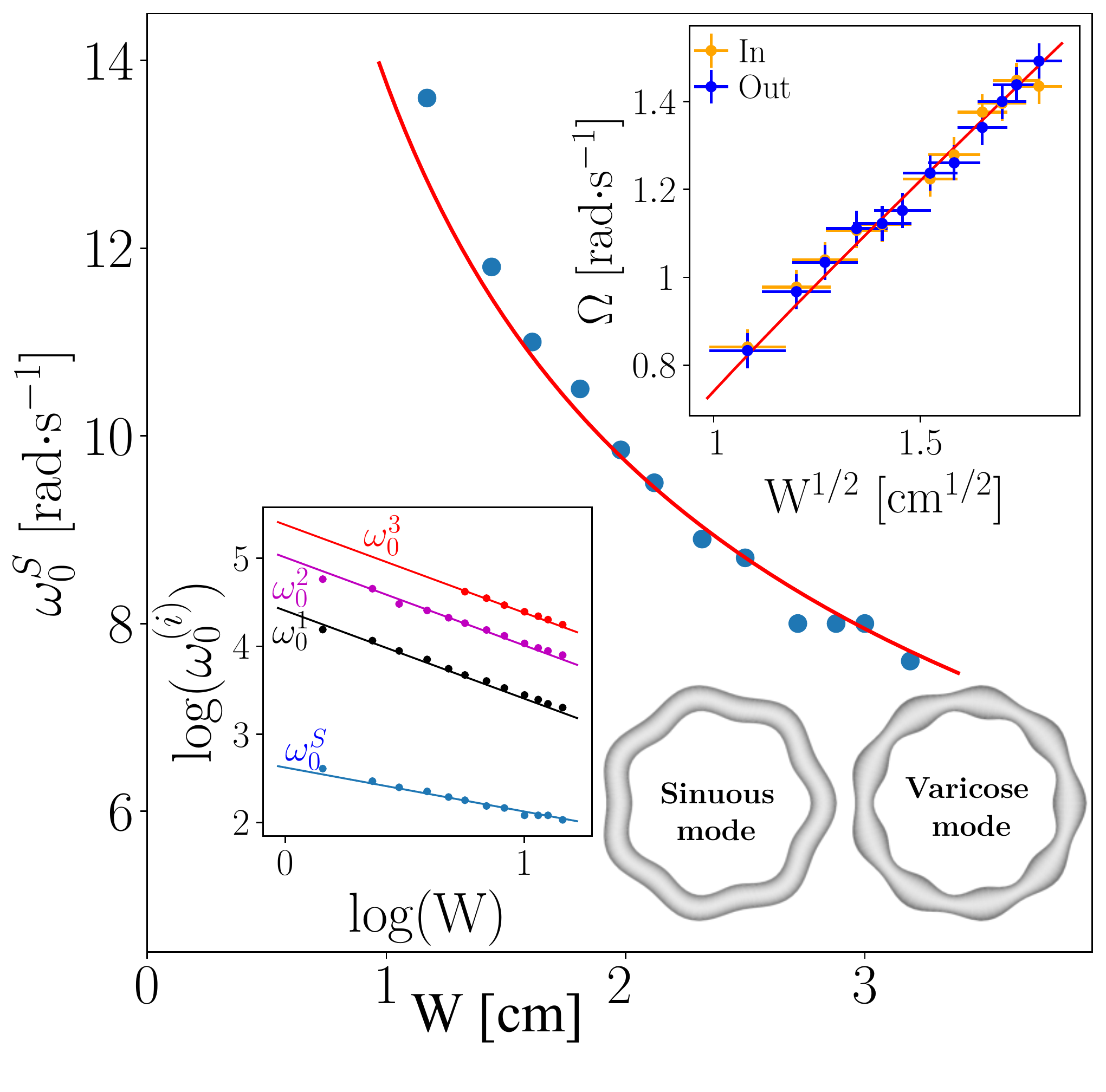}
\caption{Cutoff frequency $\omega_{0}^{\textit{S}}$ of the sinuous \textit{S} branch versus the torus width $W$. Solid line displays Eq. \eqref{eq:cutoff} with no fitting parameter. Bottom inset: Log-log plot of the cutoff frequencies versus $W$ for sinuous and sloshing branches: $\omega_0^{\textit{S}}\propto W^{-1/2}$ for the \textit{S} branch, and $\omega_0^{(i)}\propto W^{-1}$ for all \textit{Sl$_i$} sloshing branches. Top inset: Angular phase velocity $\Omega$ of the outer (blue) and inner (orange) borders for different $W$ (inferred from the \textit{V} branch). The solid line corresponds to Eq. \eqref{eq:velocity}. Insets: Sinuous and varicose mode schemes for a slim torus.}
\label{fig04}
\end{figure}

By changing the torus volume, we infer $\Omega_\varphi \equiv [\omega_V(k_\theta)/k_\theta]|_{k_{\theta}=0}$ as a function of the width $W$, on the inner and outer borders separately, as shown in the top inset of Fig. \ref{fig04}. $\Omega_\varphi$ is experimentally found to match for both borders at a given volume, and to scale as $W^{1/2}$. It is well described (see solid line in top inset of Fig. \ref{fig04}) by
\begin{align}\label{eq:velocity}
\Omega_\varphi=\frac{\sqrt{g\eff W/2}}{R_c}
\end{align}
with no fitting parameter, in agreement with Eq.~\eqref{eq:dispersion-relation} when $k_\theta\rightarrow 0$.

\paragraph*{S mode.\textemdash}The \textit{S} branch has a very different behavior, intersecting the upper sloshing branches (see Fig. \ref{fig03}). The \textit{S} branch is found to display a pure capillary regime (i.e., $\omega^2\sim k_\theta^3$) for $k_\theta \gg 0$, and to be well described (see dashed line in Fig.~\ref{fig03}) by
\begin{align}
\omega_{S}^2=\left(\omega^{S}_0\right)^{2}+\frac{\sigma\eff}{\rho} \left(\frac{k_\theta}{R_i}\right)^3.
\label{Smode}
\end{align}
with no fitting parameter, $\omega^{S}_0$ being the cutoff frequency. The \textit{S} mode being more significant for the inner border (see Fig.~\ref{fig02}), it is thus consistent that Eq.\ (\ref{Smode}) involves the inner radius $R_i$ as a length scale. Moreover, $\omega^{S}_0$ is experimentally found to be dependent on the torus width as $W^{-1/2}$ (see main Fig. \ref{fig04} and bottom inset). Since this cutoff corresponds to the motion of the center of mass of the torus section, we model it as (see Supp. Mat. \cite{SuppMat})
\begin{align}\label{eq:cutoff}
\omega^{S}_0=\frac{\pi}{2}\sqrt{\frac{g\sin(2\alpha)}{2W}}
\end{align}
Eq. \eqref{eq:cutoff} is found in good agreement with the data with no fitting parameter (see dashed line in main Fig. \ref{fig04} and bottom inset).

\paragraph*{Sloshing modes.\textemdash}The \textit{Sl$_i$} branches in Fig.~\ref{fig03} are ascribed to sloshing modes driven by gravity. These branches are well described (see dot-dashed lines in Fig.~\ref{fig03}) by
\begin{align}
\omega^2_\textit{Sl$_i$}=\left(\omega_0^{(i)}\right)^2+c^2\left(\frac{k_\theta}{R_c}\right)^2,
\label{Slmode}
\end{align}
with $c=\sqrt{g\eff R_c}$, and $\omega^{(i)}_0$ the cutoff frequencies at $k_\theta=0$. 
These nondispersive sloshing modes ($\omega^2\sim k_\theta^2$) for $k_\theta \gg 0$ differ from theoretical results on sloshing in toroidal containers \cite{Ibrahim2005} or in non-rectangular ones \cite{groves_1995}. Moreover, the cutoffs are found to scale as $\omega^{(i)}_0 \sim W^{-1}$ (see bottom inset of Fig.~\ref{fig04}), and are well described (see solid lines) by  $\omega_0^{(i)}=M(i)\sqrt{g\eff R_c}/W$ with $M(i)$ a function computed numerically (see Supp. Mat. \cite{SuppMat}).

\paragraph*{Thin torus.\textemdash}For slim tori, the inner and outer border motions interact. The \textit{S} and \textit{V} branches are then visible both on the inner and outer border spectra (see Supp. Mat. \cite{SuppMat}). The \textit{V} branch corresponds to a varicose mode, i.e., waves propagating along both borders in anti-phase motion (see inset of Fig. \ref{fig04}). The \textit{S} branch is related to a sinuous mode (also called zigzag mode \cite{PucciJFM2013,PototskyPRF2016}) for which waves on both borders are in-phase (see inset of Fig. \ref{fig04}). Indeed, by summing (resp. subtracting) signals of the outer and inner borders and taking the spectrum of the resulting signal, the sinuous (resp. varicose) branch in the total spectrum is removed, as expected for in-phase (resp. anti-phase) motions. For wide tori, interaction between the two borders vanishes for almost all $k_{\theta}$, and these modes are no longer phase related. The wide torus criterion is $W>2R_c^2/(k_\theta R_o)$, from Eq.~(\ref{eq:dispersion-relation}).

\paragraph*{Polygons.\textemdash} We now force the torus at a fixed frequency, $f$, to generate standing azimuthal waves along both borders of the torus. The torus boundary condition being periodic, the wave dispersion relation is quantized ($k_\theta$ is an integer). This quantization is studied more closely here, as well as the coupling between the outer and inner torus borders.  

\begin{figure}[t!]
 \includegraphics[width=0.96\columnwidth]{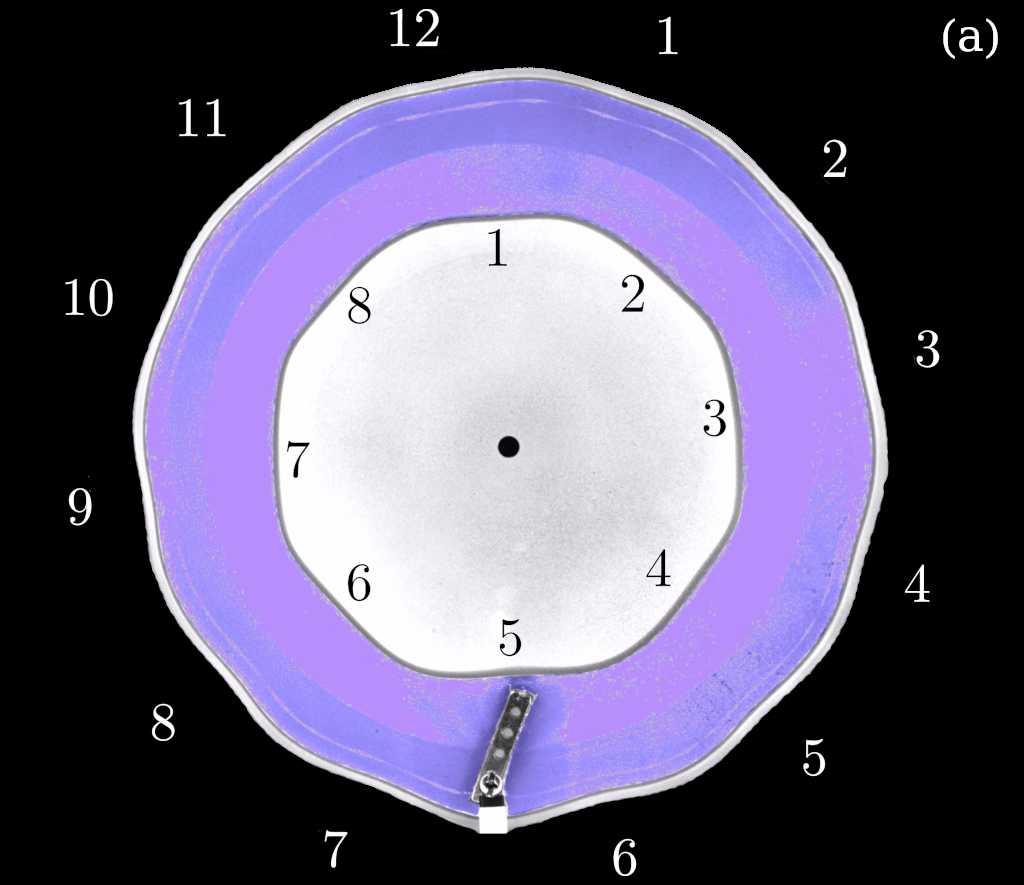}
 \includegraphics[width=\columnwidth]{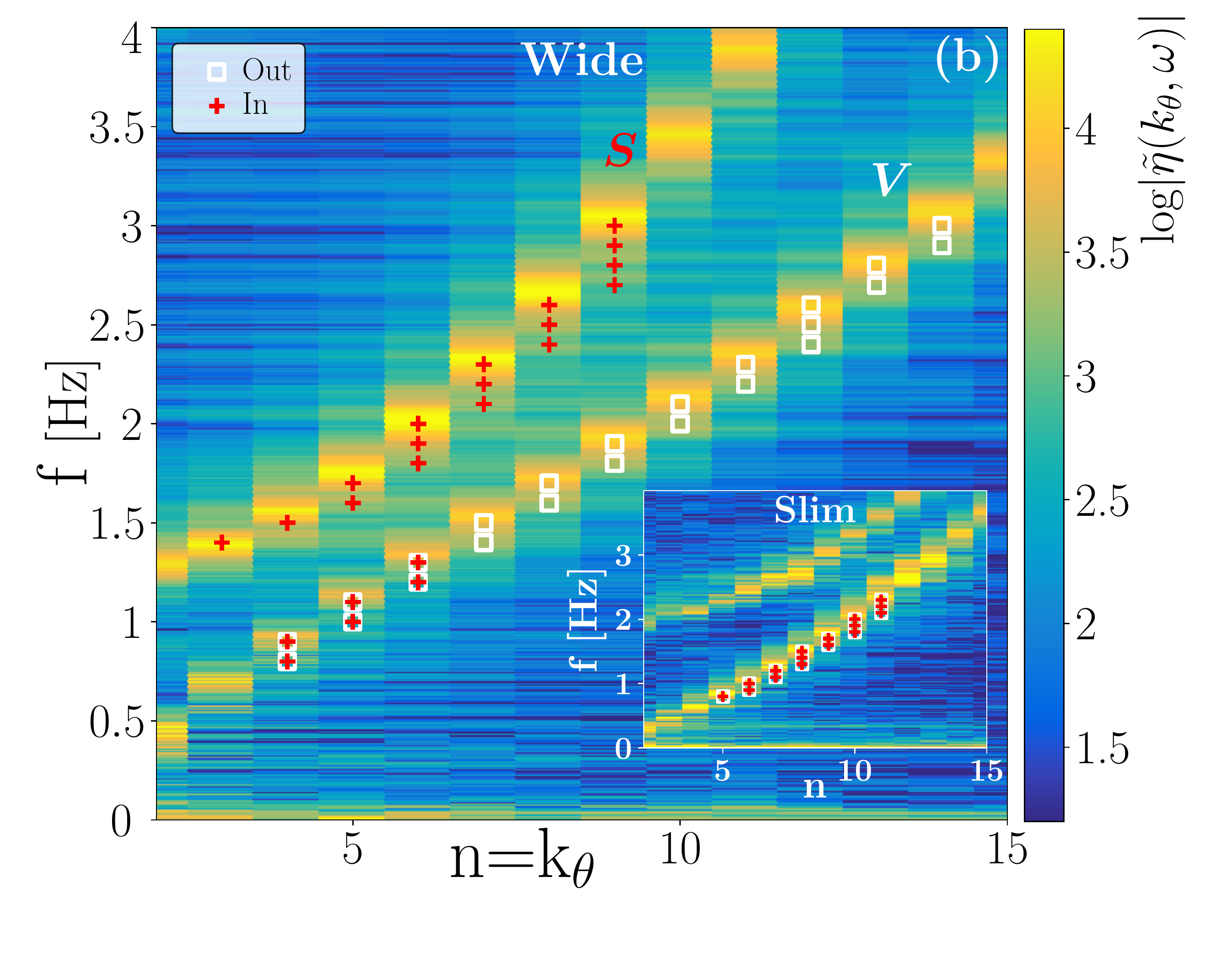}
\caption{(a) Polygonal patterns generated as standing waves along the inner border $n=8$ (octagon), and the outer border  $n=12$ (dodecagon). Forcing frequency $f=2.5$\,Hz. $W=3$ cm. Torus is colored in magenta. (b) Number of sides $n$ of the polygonal pattern on the outer ($\square$) and inner ($+$) border of a wide torus ($W=3$\,cm) for different $f$, together with the superimposed dispersion relations. For $f\geq 1.2$ Hz, the two borders decouple and $n$ differs on the two borders: the outer $n$ keeps following the \textit{V} branch, while the inner $n$ switches to the \textit{S} branch. Inset: Same for a slim torus ($W=1.5$ cm): $n$ on both borders matches for all $f$.}
\label{fig05}
\end{figure}

We consider two different torus volumes, a slim one ($W=1.5$ cm), and a wider one ($W=3$ cm). For each $f$, a short video of the torus is taken (see Supp. Mat. \cite{SuppMat}). Once the standing wave regime is established, a polygonal pattern is observed (see inset of Fig. \ref{fig05}a). We then measure the number $n$ of sides of the pattern. By varying $f\in[0.8{\rm ,\ }3]$ Hz, we plot $n$ as a function of $f$. In the slim case, $n$ is found to be the same for both the inner and outer borders within this frequency range (see inset of Fig. \ref{fig05}b). The maximal number of sides that is possible to observe can be predicted. It occurs when the wavelength is of the order of the effective capillary length, i.e., $\lambda=\ellcap=\sqrt{\sigma\eff/(\rho g\eff)} \approx 8$ mm. For $R_o=8.5$ cm, the maximal number is given by $n_m=2\pi R_o/\ellcap=67$. Note that such azimuthal patterns were previously observed up to $n=25$, but only along the outer border of a torus \cite{FalconPRL2019}.

The measurements are repeated for a wider torus where the interaction between both borders is weak. Figure \ref{fig05}b shows the superimposition of the outer and inner border wave spectra together with the number of sides $n$ for a given forcing frequency $f$. For both borders, they are the same and follow the \textit{V} branch, up to a given frequency ($f \leq1.2$ Hz). Once $f$ is large enough, $n$ becomes different on the two borders: the number $n$ for the outer border ($\square$-symbol) keeps on following the  \textit{V}-branch, while for the inner border ($+$-symbol), $n$ switches to the upper  \textit{S}-branch. This effect is due to a decoupling between the two interfaces. For the slim torus, energy on the inner and outer borders is distributed on both \textit{S} and \textit{V} branches (see Supp. Mat. \cite{SuppMat}), and $n$ matches only the \textit{V} branch as a varicose mode (see inset of Fig.~\ref{fig05}b). For the wide torus, the inner border localizes its energy much more on the \textit{S} branch (see Figs.~\ref{fig02}a) whereas the  energy on the outer border is mainly distributed on the \textit{V} branch (see Figs.~\ref{fig02}b). Except for large scales, i.e., $k_\theta \leq 2R_c^2/(WR_i)$ or $k_\theta \leq  6$ for $W=3$ cm, the two borders are decoupled (see main Fig.~\ref{fig05}b).

\paragraph*{Conclusion.\textemdash}We developed an original technique to create a stable torus of fluid moving almost completely unconstrained, and offering a system with atypical boundary conditions, i.e., periodic, curved and one-dimensional. We reported direct measurements of the dispersion relation of azimuthal waves propagating along the boundaries of this stable torus of fluid. The dispersion relation is quantized since the periodicity induces a wave number selection for a given forcing frequency band. The wave energy distribution in Fourier space is complex and exhibits several modes modeled as varicose, sinuous or sloshing ones.  In the future, this new system could highlight nonlinear phenomena such as nonlinear waves and solitons in curved geometry. It could also address the role of finite-system size effects in wave turbulence \cite{FalconDCDS2010,FalconARFM2022,HrabskiPRE2020}, as well as the coupling of waves propagating on two curved interfaces \cite{IssenmannEPL2016,PototskyPRF2016}.

\begin{acknowledgments}
We thank A. Di Palma and Y. Le Goas for their technical help on the experimental setup. Part of this work was supported by the French National Research Agency (ANR DYSTURB project No. ANR-17-CE30-0004), and by a grant from the Simons Foundation MPS N$^{\rm o}$651463.
\end{acknowledgments}


\end{document}